\documentclass[a4paper,12pt]{article}
\usepackage{amsmath, latexsym, amsfonts, amssymb, amsthm, amscd}
\usepackage{amssymb,amsthm,latexsym}

\newcommand{\tp}{{\tilde{p}}}

\newtheorem{theorem}{Theorem}

\newtheorem{lemma}{Lemma}
 \title{A field-theoretical approach to the spin glass transition: \\
models with long but finite interaction range}

\author{
Silvio Franz\footnote{\
 e-mail: {\tt franz@ictp.trieste.it}} \\
{\small {\itshape The Abdus Salam
International Center for Theoretical Physics, Condensed Matter Group}}\\
{\small {\itshape     Strada Costiera 11, P.O. Box 586, I-34100 Trieste, Italy}}\\
Fabio Lucio Toninelli\footnote{\
e-mail: {\tt fltonine@ens-lyon.fr}}\\
{\small {\itshape
Laboratoire de Physique, UMR-CNRS 5672, ENS Lyon, 46 All\'ee d'Italie, 69364 Lyon Cedex 07, France}}\\
{\small {\itshape Presently on leave at: Institut f\"ur Mathematik, Universit\"at Z\"urich
}}}

 \date{\today} 

\begin{document}

 \maketitle

\begin{abstract}
  We study spin glasses with Kac type interaction potential for small
  but finite inverse interaction range $\gamma$. Using the theoretical
  setup of coupled replicas, through the replica method we argue
  that the probability of overlap profiles can be expressed for small $\gamma$
  through a large-deviation functional. This result is supported by rigorous arguments,
showing that the large-deviation functional provides at least upper bounds for the 
probability.
Finally we analyze the rate function, in the vicinity of the critical point $T_c=1$, $h=0$
of mean field theory, and we study the free energy cost of overlap interfaces, assuming 
the validity of a gradient expansion for the rate functional.

\end{abstract}

\section{Introduction}

In recent times the introduction of interpolating techniques by
Guerra \cite{broken}, and their smart generalization by Talagrand
\cite{talaparisi} to the case of two coupled replicas of the system,
has led to the much awaited proof of the Parisi Ansatz \cite{MPV} for mean field
spin glasses.  The nature of the glassy phase of disordered systems in
finite dimension remains on the other hand a largely open problem.

The question is relevant both for the spin-glass problem, modeled by
Hamiltonians with disordered two-body interactions, and for the
structural glass problem, for which, despite the absence of intrinsic
quenched disorder, a phenomenological analogy has been suggested 
with the so called $p$-spin models, i.e., mean-field spin-glasses with
$p$-spin ($p>2$) random interactions. 

In both contexts, replica theory suggests that the mean field
equilibrium phase diagram and the nature of the glassy phase should
remain qualitatively unchanged above a finite critical dimension
\cite{dedoko}.  This description has been thwarted by phenomenological
descriptions -- droplet theories \cite{droplets} -- arguing that
Replica Symmetry Breaking is unstable in any finite-dimensional space.
Droplet theories have been comforted by mixtures of rigorous and
heuristic arguments which, while clarifying specific aspects of
finite-dimensional disordered systems, do not give a definite answer
to the question of the validity of the mean field phase diagram in
finite dimension. In three-dimensional systems it is unlikely that
the question can be solved experimentally or by numerical
simulations. The dynamics of experimental spin glasses displays
features such as a strong aging effect \cite{age} and a non-trivial
fluctuation-dissipation ratio \cite{ocio} as predicted by mean field
theory \cite{dyn}. However, the observed dynamical regime is too far
from the asymptotic situation that would allow to infer properties of
the equilibrium phase \cite{FMPP}. In simulations, on the other hand,
while dynamics suffers from the same problems as the experiments, the
possibility of studying equilibrated samples is restricted to small
systems and it cannot be excluded that the observed mean-field like
behavior is dominated by finite-size effects.  In this situation it is
natural to look at models which, interpolating between mean field and
finite range, allow for analytical treatment. The utility of such
models would not be confined to the study of the equilibrium
phase. Neglection of space in the mean-field dynamics of the $p$-spin
model, for example, predicts an artifactual dynamical breaking of
ergodicity that blocks the system to free-energy densities higher then
the equilibrium one, realizing the scenario predicted by the idealized
Mode-Coupling Theory for structural glasses \cite{MCT}. The inclusion
of finite-dimensional effects is necessary to study the dynamical
processes that restore ergodicity allowing to go beyond Mode-Coupling
Theory.

In recent work \cite{Kaclim} we showed that,
considering spin glasses with variable range of interaction
$\gamma^{-1}$ and taking the Kac limit of vanishing $\gamma$, one
recovers the mean field free energy, thus generalizing the celebrated
Lebowitz-Penrose theorem \cite{lp}. In addition to the free energy, the same
convergence result applies to the local order parameter
\cite{KacII}. In other words, one can define a local overlap distribution function and
show that this tends to the usual mean field overlap distribution
function $P(q)$, which is non-trivial in the low temperature
regime. Of course, this has no direct implication on the phase
structure for small but finite $\gamma$, where one can expect a
transition only for high enough dimensionality $d$. One can hope
however that the $\gamma=0$ limit can be taken as a starting point to
study the spin glass transition in a simplified setting. This hope is
comforted by what happens in non-disordered systems, e.g., the
ferromagnetic Ising model. In that case, through a  block-spin
transformation one can introduce spatial magnetization profiles and
express their probability by means of a suitable free-energy
functional \cite{abcp}. This leads to a field theory over the magnetization
profiles, which can be analyzed ``semiclassically'', the role of the
large parameter in front of the action being played by
$\gamma^{-d}$. Thanks to that, customary low temperature techniques,
of the kind of the Peierls argument or Pirogov-Sinai theory for Ising
systems, can be adapted to prove the existence of a phase transition
for small $\gamma$ in dimension large enough \cite{cp}.

In this paper
we address the problem of achieving an analogous field-theoretical
representation for the spin glass problem. 
We consider a variant of the spin glass model with Kac interaction
potential, the ultimate goal being that of studying the presence of long-range order for
large but finite interaction range $\gamma^{-1}$.
We
 first give a  definition of long-range order as
sensitiveness in the bulk to appropriately chosen boundary
conditions. We apply to our problem the method of coupled replicas
\cite{fpvI} and, as in \cite{fpvII}, we consider two copies of the same
system coupled in such a way to have local overlap fixed on the
boundary.  The appropriate field theory will then result from the
evaluation of the free-energy cost for having overlap profiles on
certain coarse-grained regions of space. 
In a first moment we 
resort to  the replica method and use a generalization of the
Parisi Ansatz suitable to study problems of coupled replicas
\cite{fpvI}, in order to evaluate the free-energy cost.  Then, through the by now usual interpolating techniques
\cite{limterm}, \cite{broken}, \cite{ass} we prove that the replica expression provides a
free-energy lower bound. 

Once the field-theoretical representation has been obtained, one can
envisage to use it to prove the existence of long-range order in high
enough dimension.  A necessary preliminary task, which we undertake in the present work, consists in the
analysis of the properties of the rate function and in the evaluation
of the cost of overlap profiles. This functional has a zero mode
corresponding to flat profiles with overlap chosen in the support of
the mean field $P(q)$ function. This implies that the lowest cost interfaces
separating regions with homogeneous different values of the overlap
will be extended in space. 

Close to the critical point $T=1$, $h=0$ of mean-field theory, we give some interface estimates
in the lines of \cite{fpv}, indicating that the cost for extended
``overlap interfaces'' grows with the linear dimension $L$ of the system
as $L^{d-5/2}/\gamma^{5/2}$. 
 The main difficulty in inferring from this fact the presence of long-range 
order for $d>5/2$ lies in the possible
existence, that we cannot exclude, of entropic contributions scaling
with higher powers of $L$, that would destroy the possibility of the
phase transition for any $\gamma>0$. 

\section{The model}

The Hamiltonian of the Kac spin glass with two-body interactions, on the  $d$-dimensional
hypercube $\Lambda=\{1,\cdots,L\}^d$ and in presence of a magnetic
field $h$, can be defined as
\begin{eqnarray}
\label{ham}
H_\Lambda(\sigma,h;J)=-\frac1{\sqrt 2}\sum_{i,j\in \Lambda}
J_{ij}\sigma_i \sigma_j-h\sum_{i\in\Lambda}\sigma_i.
\end{eqnarray}
The $J_{ij}$ are Gaussian independent random variables with zero mean
and variance
\begin{eqnarray}
E J_{ij}^2=\gamma^d \phi(\gamma|i-j|),
\end{eqnarray}
where $\phi(|x|)$ is a smooth non-negative function with 
support in  $|x|\le1$, and normalized so that
\begin{equation}
\int d^d x\,\phi(|x|)=1.
\end{equation}
We will also require $\phi$ to be non-negative definite, i.e., 
\begin{equation}
\label{nonneg}
\int d^d x\,\phi(|x|)e^{i k x}\ge0 \;\;\mbox{for any\;\;} k\in{\mathbb R}^d.
\end{equation}
In the theory of mean field spin glasses an important role is
played by $p$-spin models where the spins are coupled through $p$-body
interactions. It is interesting to generalize the definition of the
Kac spin glass model to that case and define the Hamiltonian
\begin{equation}
\label{hP}
H_\Lambda^{(p)}(\sigma,h;J)=-\frac1{\sqrt{2}}\sum_{i_1,\cdots, i_{p}\in\Lambda }
J_{i_1\cdots i_{p}}
\sigma_{i_1}\cdots\sigma_{i_{p}}
-h\sum_{i\in\Lambda}\sigma_i,
\end{equation}
with 
\begin{eqnarray}
\label{wij}
E(J_{i_1\cdots i_{p}}^2)=
\frac{\sum_{k\in \Lambda}\psi(\gamma|i_1-k|)\cdots
\psi(\gamma|i_p-k|)}{W(\gamma)^{p}}
\end{eqnarray}
and
\begin{equation}
W(\gamma)=\sum_{k\in \Lambda}\psi(\gamma|k|). 
\end{equation}
where $\psi(|x|)$, $x\in {\mathbb R}^d$, is any non-negative summable
function with compact support, normalized as
\begin{equation}
\int d^dx\, \psi(|x|)=1.
\end{equation}
The rationale for the choice (\ref{wij}) among all possible functions
of $p$ variables with range $\gamma^{-1}$ is discussed in \cite{KacII}
and ensures that for all $\beta,h,\Lambda$ and for even $p$, the finite
volume free-energy of the Kac model is bounded  below by the mean
field free-energy corresponding to the same external parameters and number of spins.
Of course terms with different $p$ can be combined.  For these
models we recently proved the convergence of the free energy to the
mean field value in the Kac limit, provided that $p$ is even, as well
as the convergence of the distribution of the local order parameter
\cite{KacII}.

For many purposes the analysis of the $p$-spin model is very similar
to the one of the usual two-body Hamiltonian (\ref{ham}), while the notations for
generic $p$ are much heavier. For that reason we will present all our
arguments in the case of the two-body model, and we will give at the end
the general formulae in the $p$-spin case.

\section{Overlap profiles and boundary conditions}

At mean field level, the order parameter for the spin glass transition is
the Parisi probability distribution, describing the statistics of the
overlap between two replicas with the same disorder, induced by the
Gibbs measure and by the disorder distribution. In Ref. \cite{KacII}
we showed that in Kac spin glasses the distribution of {\it
local} overlaps, on scales of the order of the interaction range, 
tends to the Parisi overlap probability distribution for
$\gamma\to0$, in any space dimension $d$. This does not tell anything about the possibility of
long-range order where the probability distribution of the global
overlap has a non-trivial shape. 
In order to discuss the possibility of long-range order, let us introduce the
local overlaps, and the probability of local overlap profiles. 
First of all, let us partition ${\mathbb Z}^d$ into cubes $\Omega_k$, 
$k\in {\mathbb Z}^d$, of side
$\delta/\gamma$ (to be chosen to be an integer):
\begin{eqnarray}
\Omega_k=\{i\in {\mathbb Z}^d\;:\;\frac{\delta}{\gamma}k_l\leq i_l<
  \frac{\delta}{\gamma}(k_l+1); \;\; l=1,...,d\}.
\end{eqnarray}
The situation we have in mind is 
\begin{equation}
\label{situation}
1\ll \frac \delta\gamma\ll \frac 1\gamma\ll L
\end{equation}
and, for definiteness, one can think that $\delta\sim \gamma^{1-\varepsilon}$ for some $0<\varepsilon<1$.
Let us denote by $M=\left(L\frac{\gamma}{\delta}\right)^d$ the total
number of boxes. Given
two spin configurations ${ \sigma^1}$ and ${\sigma^2}$ on $\Lambda$
we define the local overlaps on the box $\Omega_k$ as
\begin{eqnarray}
q_k({\sigma^1},{\sigma^2})=\left(\frac{\gamma}{\delta}\right)^d
\sum_{i\in\Omega_k}\sigma_i^1\sigma_i^2
\end{eqnarray}
and the global overlap 
\begin{eqnarray}
q({\sigma^1},{ \sigma^2})=\frac{1}{L^d}
\sum_{i\in\Lambda}\sigma_i^2\sigma_i^2
\end{eqnarray}
Of
course, as long as $\delta/\gamma$ is finite the local overlap can
assume just a finite number of values.
We define a ``local overlap profile'' $\{\tp_k\}_{k=1,\cdots,M}$ to be a
collection of possible values for the local overlaps\footnote{The definition we 
give here does not coincide exactly with
the one adopted in \cite{KacII}, while its properties for small
$\gamma$ do not depend on the details of the definition}.  Given a
realization ${J}$ of the disorder we define the Boltzmann weight
associated to a local overlap profile $\{\tp_k\}$ as
\begin{eqnarray}
B_\Lambda[\{\tp_k\};J]=\sum_{\sigma^1,\sigma^2} \exp(-\beta H_\Lambda
(\sigma^1,h;J)-\beta H_\Lambda(\sigma^2,h;J))1_{\{q_k( \sigma^1,\sigma^2) =\tp_k 
\forall k\}}.
\end{eqnarray}
For fixed disorder, the probability of an overlap
profile is just
\begin{eqnarray}
P_\Lambda[\{\tp_k\};J]=\frac{B_{\Lambda}[\{\tp_k\};J]}
{Z_{\Lambda}(\beta,h;J)^2}. 
\end{eqnarray}
From the results of \cite{KacII}, one can argue that the functional
$P_\Lambda[\{\tp_k\};J]$ must be concentrated, in the situation
(\ref{situation}), on profiles where each of the $\tp_k$ is in
$Supp(P(q))$, the support of the mean field Parisi overlap
distribution.  The expectation from replica theory is that in high
enough dimension and $\gamma$ small enough, at temperatures smaller
than $T_c=1$, long-range order should be present.  The two-replica
free energy would be dominated by configurations with a constant
overlap in space, their weight being given by the function $P(q)$.

The space homogeneity of typical overlap profiles can be argued considering that in any fixed subset
$\Omega\subset \Lambda$ containing an extensive number of spins
$\alpha |\Lambda|$ the overlap probability function restricted to this
set should, in the limit of large $\Lambda$, coincide with the one of
the whole system. Considering then the total overlap between two
configurations $q=\alpha q_\Omega+(1-\alpha)q_{\Lambda-\Omega}$, it is
clear that from the fact that $q_\Omega^2$ and $q_{\Lambda-\Omega}^2$
are distributed like $q^2$ it follows that the product $q_\Omega
q_{\Lambda-\Omega}$ is also distributed like $q^2$. Analogously for all
$r$ and $s$ positive integers, $q_\Omega^r q_{\Lambda-\Omega}^s$ must be
distributed like $q^{r+s}$, implying that
$P(q_\Omega|q_{\Lambda-\Omega})=\delta(q_\Omega-q_{\Lambda-\Omega})$,
i.e. homogeneity of the overlap in space.\footnote{We thank J. Kurchan
for discussions on this point.} 

In order to detect the possibility of
an ordering phase transitions one can then impose ``overlap boundary
conditions'' in the two-replica system.  Suppose to fix the overlaps
$\tp_k$ to a given value $\tp\in Supp(P(q))$ for all the $\Omega_k$
belonging to a boundary region of thickness $\sim \gamma^{-1}$ around
the border, and to let otherwise free the boundary
conditions. Long-range order would correspond to the fact that the
probability of having a deviation of the local overlap from the value
$\tp$, in a box situated in the bulk of the system, is vanishing for
diverging box size.

In the following, we will concentrate on estimates of
\begin{eqnarray}
\label{freeen}
F_\Lambda[\{\tp_k\}]=-\frac{1}{\beta}E\,\log
B_\Lambda[\{\tp_k\};J],
\label{free}
\end{eqnarray} 
which is the  disorder-averaged free energy functional corresponding to $B_\Lambda[\{\tp_k\};J]$,
i.e., the free energy of two replicas with fixed overlap profile. 
We will see in the next sections 
that  saddle point approximation can be used to obtain a small-$\gamma$ 
expansion of the free energy functional.
 
The rationale behind the definition (\ref{freeen}) is that we expect 
the fluctuations of the probability $P_\Lambda[\{\tp_k\};J]$ to 
be much smaller than its typical value, as long as $P_\Lambda\sim \exp(-a L^\mu)$ for some
$a,\mu>0$, so that the expectation of the logarithm gives its typical 
value\footnote{It is easy to see, by the means
of the usual interpolation techniques, that the fluctuations of $\log P$
cannot be larger than $O(L^{d/2})$, which implies the desired self-averaging
property for those profiles such that $\mu > d/2$.}.
Thanks to the  the same self-averaging argument, 
it is reasonable to expect that the free energy of the two-replica system with fixed overlap
boundary conditions can be obtained just considering the minimum of
$F_\Lambda[\{\tp_k\}]$ over
all the overlap profiles $\tp_k$ satisfying the boundary conditions. 
This (unproven) self-averaging property will be implicitly assumed to hold in the rest of the paper.

\section{Replica analysis}

The free energy $F_\Lambda[\{\tp_k\}]$ can be evaluated through the replica
method for a system of constrained copies, which shows in a natural way
how $F_\Lambda[\{\tp_k\}]$ takes, in the large-volume and small-$\gamma$ 
limit, the form of a rate function in a large
deviation problem. For the sake of briefness we will not
reproduce the full calculations, and we will just indicate the main steps in
the derivation and the final results. In the next section we will then
prove that, modulo some error terms which we
will estimate, the rate functional found via the replica method provides a
rigorous free energy lower bound, as it happens in the case of the
unconstrained model \cite{broken}.

As usual, the replica method starts from the
expression of $E(B_\Lambda[\{\tp_k\};J]^n)$ for integer $n$. 
Introducing the replicas $\sigma_i^{r,a}$ for $r=1,2$ and $a=1,...,n$ 
and the local replica matrices 
$Q_k^{ra,sb}=\left(\frac{\gamma}{\delta}\right)^d \sum_{i\in
    \Omega_k} \sigma_i^{r,a}\sigma_i^{s,b}$
satisfying the symmetry conditions 
$Q_k^{ra,sb}=Q_k^{sb,ra}$
and such that the diagonal elements are fixed to 
$Q_k^{ra,ra}=1$
and
$Q_k^{1a,2a}=\tp_k$, 
the average of the $n$-th moment
can be written as an integral over the
elements $Q_k^{ra,sb}$ which are not fixed:
\begin{eqnarray}
\label{mom_n}
E(B_\Lambda[\{\tp_k\};J]^n)&=&
\int {\cal D}Q \exp\left(\frac{\beta^2}{4}
\left( \frac{\delta}{\gamma}\right)^d \sum_{k,m}\delta^d
\phi(\delta|k-m|)
\sum_{a,b}^{1,n}\sum_{r,s}^{1,2}  Q_k^{ra,sb} Q_m^{ra,sb}\right)
\nonumber\\
&&\times
\sum_{\{{\bf \sigma}\}}
e^{\beta h\sum_{i,r,a}\sigma_i^{r,a}}
1_{\{q_k({\bf \sigma}^{r,a},{\bf \sigma}^{s,b})=Q_k^{ra,sb};\; \forall \;
  a,b,r,s,k\}}.\nonumber
\end{eqnarray}
This expression is simplified introducing a continuum notation. Let
us rescale the system by a factor $\gamma$ and let $x\in V\equiv
[0,L\gamma]^d$. Define then $Q_x^{ra,sb}=Q_k^{ra,sb}$ for $x/\gamma\in
\Omega_k$ and write
\begin{eqnarray}
E(B_\Lambda[\{\tp_k\};J]^n)
&=&\int{\cal D}Q \exp\left(-\beta
F^{Kac}_{(n)}[Q,\{\tp_k\}]\right) \\\nonumber &\equiv&
\int{\cal D}Q
\exp\left(-\frac{\beta^2}{8\gamma^d} \sum_{r,s,a,b} \int_V
d^dx\, d^dy\, \phi(|x-y|)(Q_x^{ra,sb}-Q_y^{ra,sb})^2\right)
\nonumber\\ &&\times\exp\left(-\frac{\beta}{\gamma^d}\int_V d^dx\,
F^{(n)}_{(\delta/\gamma)^d} [Q_x,\tilde p_x]\right)
\end{eqnarray}
where $F^{(n)}_{(\delta/\gamma)^d}$ is the free energy functional
which one finds, within the replica method, when one computes the free
energy of two Sherrington-Kirkpatrick (SK) systems with $(\delta/\gamma)^d$ spins, constrained
to have a mutual overlap $\tp_x$. Note that $F^{(n)}$ has just a local dependence on the
``overlap matrix'' $Q_x$ and on the overlap profile $\tp_x$.
For large $\delta/\gamma$ a saddle
 point approximation can be considered and one needs 
an Ansatz for the matrix form in order to compute the $n\to 0$ continuation. 
This has been considered in \cite{fpvI,fpvII,fp} where the problem was 
treated assuming that each $n\times n$ matrix $\{Q_x^{ra,sb}\}_{1\le
a,b\le n}$, for fixed indexes $r,s$, has a Parisi-like structure
\cite{MPV}, and is therefore parametrized, in the limit $n\to0$, by a
functional order parameter
$
q^{r,s}_x(.):[0,1]\longrightarrow [0,1],
$ 
verifying the symmetry $q^{r,s}_x(u)=q^{s,r}_x(u)$ and the
monotonicity condition that for $0\leq v\leq u\leq 1$, the $2\times 2$
matrix in the indexes $r$ and $s$,
$\{q_x^{r,s}(u)-q_x^{r,s}(v)\}_{1\le r,s\le 2}$ are non-negative
definite. One therefore needs to consider two families of functions
$q_x(u)=q^{1,1}_x(u)=q^{2,2}_x(u)$ and $p_x(u)=q^{1,2}_x(u)=q^{2,1}_x(u)$, with
$x\in V$ and $u\in[0,1]$.

\subsection{Analytic continuation}

In order to perform the analytic continuation of $F^{Kac}_{(n)}$ as $n\to0$,
it is 
useful to define the convolution of the functions $q_x^{r,s}(u)$
with the interaction potential 
\begin{equation}
\hat q^{r,s}_x(u)=\int_V d^d y\,\phi(|x-y|)q^{r,s}_y(u),
\end{equation}
and the Lagrange multiplier $\epsilon_x$ associated to the 
constrained local overlaps $\tp_x$. 
After some algebra  one can write for the $n\to0$ limit:
\begin{eqnarray}
-\beta \gamma^d F^{Kac}_\Lambda[\{\tp_x\},\{q_x^{r,s}(.)\},\{\epsilon_x\}]= (2\log 2)|V|
-\frac{\beta^2}{2}\int_V d^d x\left(1+\tp_x^2-\int_0^1 du\,
\left(q_x(u)^2+p_x(u)^2\right)\right)\\
\label{kin}
+\frac{\beta^2}{4}\int_{V\times V}d^dx\, d^dy \phi(|x-y|)
\left((\tp_x-\tp_y)^2-\int_0^1 du\,[(q_x(u)-q_y(u))^2+(p_x(u)-p_y(u))^2]
\right)\\
+\int_V d^d x \left(-\epsilon_x\tp_x+\log \cosh(\epsilon_x)
+g_x(0,h,h;\epsilon_x)\right),
\label{fgdrep2}
\end{eqnarray}
where $g_x(u,y_1,y_2;\epsilon_x) $, with $y_1,y_2\in \mathbb R$, $u\in[0,1]$, is 
 the solution 
of the 
backward parabolic equation\footnote{
One often considers the case where the functions $q^{r,s}_x(.)$ are piecewise constant. In this case,
Eq. (\ref{eq:antipara}) has to be interpreted correctly, see Section \ref{sec:int_est}}
\begin{eqnarray}
\label{eq:antipara}
\frac{\partial g_x}{\partial u}= -\frac{1}{2}\sum_{r,s}^{1,2}
\frac{\partial \hat{q}_x^{rs}}{\partial u}\left( \frac{\partial^2
g_x}{\partial y_r\partial y_s}+u \frac{\partial g_x}{\partial
y_r}\frac{\partial g_x}{\partial y_s} \right)
\end{eqnarray}
with final conditions
\begin{eqnarray}
\label{final}
g_x(1,y_1,y_2;\epsilon_x)=
\log\left[\cosh(\beta y_1)\cosh(\beta y_2)\right]
+\log \left(1+\tanh(\epsilon_x)\tanh(\beta y_1)\tanh(\beta y_2)\right).
\end{eqnarray}
The functions $q^{r,s}_x(.)$ and $\epsilon_x$ are variational parameters over which one has
to optimize to find the desired free energy (\ref{freeen}) as a function of the overlap profile.

It is instructive to compare the Kac functional with the formula of the Parisi mean field free energy
which, once optimized over $q(.)$, gives the free energy per spin of the SK model:
\begin{equation}
\label{parisi1r}
-\beta {\mathcal F}^{Parisi}[q(.)]= \log 2-\frac{\beta^2}4\left(1-\int_0^1du\, q(u)^2\right)+f(0,h),
\end{equation}
$f(u,y)$ being the solution of 
\begin{eqnarray}
\label{eq:antipara1}
\left\{
\begin{array}l
\frac{\partial f}{\partial u}= -\frac{1}{2}
\frac{d q(u)}{d u}\left( \frac{\partial^2
f}{\partial y^2}+u \left(\frac{\partial f}{\partial
y}\right)^2 \right)\\
f(1,y)=\log\cosh(\beta y)
\end{array}
\right..
\end{eqnarray}

\subsection{The $p$-spin case}

We give here without proof the expression of the Kac free-energy functional for the
$p$-spin model: 
\begin{multline}
\label{fgdreppspin}
-\beta \gamma^d F_\Lambda^{Kac,p}[\{\tp_k\},\{q_k^{r,s}(.)\},\{\epsilon_k\}]=2\log 2
|V|
-\frac{\beta^2(p-1)}{2}
\int_V d^d x_1\cdots d^d x_p\,\phi(x_1,\cdots,
x_p)\\
\times\Big[1+\tp_{x_1}\cdots\tp_{x_p}-\int_0^1 du\left(
q_{x_1}(u)\cdots q_{x_p}(u)+p_{x_1}(u)\cdots p_{x_p}(u)\right)
\Big]\\
+\int_V d^d x \left(-\epsilon_x \tp_x+\log \cosh(\epsilon_x)+\bar 
g_x(0,h,h;\epsilon_x)\right),
\end{multline}
where 
$$
\phi(x_1,\cdots,x_p)=\int_{\mathbb R^d}d^d k\,\psi(|x_1-k|)\cdots
\psi(|x_p-k|)
$$
and 
$\bar g_x(u,y_1,y_2;\epsilon_x)$ is the solution of 
the backward equation (\ref{eq:antipara}), with $\hat q^{r,s}_x(u)$ replaced by 
$$
\frac p2\int_V d^d x_2\cdots d^dx_p\, \phi(x,x_2,\cdots,x_p)
q_{x_2}^{r,s}(u)\cdots q_{x_p}^{r,s}(u)
$$
and with the same final condition (\ref{final}).

\section{Interpolating estimates}

\label{sec:int_est}

In this section show how the free energy functional (\ref{fgdrep2}) arises, without using the
replica formalism. 
We start by fixing more precisely notations and definitions.
For any box $\Omega_k$, let $q^{r,s}_k(u)$, $r,s=1,2$ be functions
$$
q^{r,s}_k(.):[0,1]\longrightarrow [0,1]
$$ 
satisfying the following conditions:
\begin{itemize}
\item symmetry:
\begin{equation}
\label{c_simm}
q^{1,1}_k(u)=q^{2,2}_k(u)\equiv q_k(u),\;\; 
q^{1,2}_k(u)=q^{2,1}_k(u)\equiv p_k(u) 
\end{equation}
\item boundary values:
\begin{equation}
\label{c_bv}
q^{r,s}_k(0)=0,\;\;q_k(1)=1,\;\;p_k(1)=\tp_k
\end{equation}
\item positive definiteness:
\begin{eqnarray}
\label{c_pd}
\mbox{for any $0\le v\le u\le1$ the matrix}
\;\;\;\{q_k^{r,s}(u)-q_k^{r,s}(v)\}_{r,s}\;\;\; 
\mbox{is non-negative definite}.
\end{eqnarray}
\end{itemize}
Then, 
\begin{theorem}
\label{th:main}
For any choice of $\{q_k^{r,s}(.)\}_k$ satisfying conditions 
(\ref{c_simm})-(\ref{c_pd}) and for any $\{\epsilon_k\}$, one has
\begin{eqnarray}
\label{b_teo}
-\beta F_\Lambda[\{\tp_k\}]\le -\beta F^{Kac}_\Lambda[\{\tp_k\},\{q_k^{r,s}(.)\},\{\epsilon_k\}]+
O(\delta)|\Lambda|.
\end{eqnarray}
\end{theorem}
One should keep in mind that $\delta$ is the size of the boxes, in the rescaled units, which 
has to be thought of as very small.\\
{\bf Remark} We expect that also a lower bound of the type (\ref{b_teo}) holds. To prove this, 
a suitable generalization of Talagrand's theorem \cite{talaparisi} 
would be needed.

{\em Proof of Theorem \ref{th:main}}
We introduce a generalization of the
Aizenman, Sims and Starr's Random Overlap Structure 
\cite{ass}, suitable to study a problem of two
coupled replicas with spatially inhomogeneous structure. 
We introduce a set of possibly random weights
$\xi_\alpha\ge0$, where $\alpha$ takes values in a discrete set of indexes, 
such that
$$
\sum_\alpha \xi_\alpha=1,
$$
and cavity fields $h_i^{\alpha,s}$, $\kappa^{\alpha,s}$, $i\in\Lambda$,
and $s=1,2$. The cavity fields are
 centered Gaussian random variables, 
independent of the couplings $J_{ij}$, with covariances
\begin{eqnarray}
\label{h}
E(h_i^{\alpha , s}h_j^{\beta , r}) &=& \delta_{ij}
\sum_{m=1}^{(L\gamma/\delta)^d} \delta^d\phi(\delta|k-m|) q_m^{\alpha,s;\beta,r}\\
\label{k}
E(\kappa^{\alpha,s}\kappa^{\beta , r}) &=&
\frac{\delta^{2d}}{2\gamma^d} \sum_{k,m=1}^{(L\gamma/\delta)^d}
\phi(\delta|k-m|) q_k^{\alpha,s;\beta,r}q_m^{\alpha,s;\beta,r},\\
E(\kappa^{\alpha,s}h_i^{\beta,r}) &=&0
\end{eqnarray}
for $i\in\Omega_k$.
We fix $q_k^{\alpha,1;\alpha,1}=q_k^{\alpha,2;\alpha,2}=1$ and
$q_k^{\alpha,1;\alpha,2}=\tp_k$, the constrained overlap in the box
$\Omega_k$, while all the other parameters are free 
(apart from the obvious constraint
that the above covariance matrices are non-negative definite) 
and can be optimized
to saturate the bounds.  Notice that the cavity fields in different
replicas are correlated. Define
$$
H_t^\alpha(\sigma^1,\sigma^2)=\sqrt{t}\sum_{s=1}^2\left(
H_\Lambda(\sigma^s,h\!=\!0;J)-\kappa^{\alpha,s}\right)-\sqrt{1-t}\sum_{i,s} h_i^{\alpha,s}
\sigma_i^s-h\sum_{i,s}\sigma^s_i,
$$ 
$Z_{\alpha,t}=Z_{\alpha,t}[\{\tp_k\}]$ as the respective partition function with 
constrained overlap profile $\{\tp_k\}$, $Z_t$ as
$$
Z_t=\sum_\alpha \xi_\alpha Z_{\alpha,t},
$$
and the interpolating free energy $\mathcal F_t$ as
\begin{eqnarray} 
-\beta \mathcal F_t=
E\log\frac{\sum_\alpha \xi_\alpha Z_{\alpha,t}}{\sum_\alpha
 \xi_\alpha e^{\beta \sum_s \kappa^{\alpha,s}}}. 
\end{eqnarray}
For  the $t$-derivative, one gets 
\begin{eqnarray} 
-\beta \frac{\partial \mathcal F_t}{\partial t}&=&
\frac{\beta^2}{4} E Z_t^{-1}
\sum_{\alpha,r,s} \xi_\alpha Z_{\alpha,t}\left( \sum_
{i,j\in\Lambda } \gamma^d \phi(\gamma|i-j|) \omega_{\alpha}
(\sigma_i^r \sigma_i^s \sigma_j^r \sigma_j^s)\right.\\\nonumber 
&&\left.+2E(\kappa^{\alpha , s}\kappa^{\alpha , r})
-2\sum_i E(h_i^{\alpha , s}h_i^{\alpha , r})  \omega_{\alpha}
(\sigma_i^r \sigma_i^s)\right)
\\\nonumber
& &
- \frac{\beta^2}{4} E Z_t^{-2}
\sum_{\alpha,\beta,r,s} \xi_\alpha \xi_\beta
Z_{\alpha,t}  Z_{\beta,t}
\left( \sum_
{i, j\in \Lambda } \gamma^d \phi(\gamma|i-j|) 
\omega^{(2)}_{\alpha,\beta}
(\sigma_i^{r,1} \sigma_i^{s,2} \sigma_j^{r,1} \sigma_j^{s,2})\right.\\\nonumber
&&\left.+
2E(\kappa^{\alpha , s}\kappa^{\beta , r})
-2\sum_i E(h_i^{\alpha , s}h_i^{\beta , r})  
\omega^{(2)}_{\alpha,\beta}
(\sigma_i^{r,1} \sigma_i^{s,2})\right).
\nonumber
\end{eqnarray}
Here, $\omega_{\alpha}(.)$ 
denotes the Gibbs average, corresponding 
to the Hamiltonian $H_t^\alpha(\sigma^1,\sigma^2)$, 
acting on the two replicas $\sigma^{1},\sigma^{2}$ with constrained
overlap profile. On the other hand, $\omega^{(2)}_{\alpha,\beta}$ is the {\em duplicated}
Gibbs average, involving the {\em four} replicas $\sigma^{r,s},r,s=1,2$, with Hamiltonian
$$
H_t^\alpha(\sigma^{1,1},\sigma^{2,1})+H_t^\beta(\sigma^{1,2},\sigma^{1,2}).
$$
Note that, in the average $\omega^{(2)}_{\alpha,\beta}$, only the overlap profiles between replicas
$\sigma^{1,1},\sigma^{2,1}$ and between $\sigma^{1,2},\sigma^{2,2}$ are constrained to $\{\tp_k\}$.
Thanks to the constraints on the local overlap profile and 
to the choice for the covariances of the cavity fields, the first term in the derivative
is at most of order $O(\delta)|\Lambda|$ 
(it is not exactly zero, since $\phi$
has small variations within each box). The second one gives
\begin{eqnarray} 
&&- \frac{\beta^2}{4} \left(\frac\delta\gamma\right)^{2d}E \left\{Z_t^{-2}
\sum_{\alpha,\beta,r,s} \xi_\alpha \xi_\beta Z_{\alpha,t} Z_{\beta,t}
 \sum_{k,m} \gamma^d\phi(\delta|k-m|)\right. \\\nonumber
&&\times\left.\omega^{(2)}_{\alpha,\beta}[
(q_k(\sigma^{r,1},\sigma^{s,2})-q^{\alpha,s;\beta,r}_k))
(q_m(\sigma^{r,1},\sigma^{s,2})-q^{\alpha,s;\beta,r}_m))]\right\}+
O(\delta)|\Lambda|.
\end{eqnarray}
The average is negative since $\phi$ is non-negative definite by the assumption (\ref{nonneg}), so that we get the
bound
\begin{equation}
-\beta F_\Lambda[\{\tp_k\}]=-\beta \mathcal F_1\le -\beta \mathcal 
F_0+O(\delta)|\Lambda|.
\end{equation}

To complete the proof of Eq. (\ref{b_teo}), we have to show that there exists a choice of the $\xi_\alpha$ 
and of the cavity fields such that $\mathcal F_0$ coincides with 
$F^{Kac}_\Lambda[\{\tp_k\},\{q_k^{r,s}(.)\},\{\epsilon_k\}]$.
As usual, it is sufficient to consider the case where the functions $q_k^{r,s}(.)$
are piecewise constant, the general case being obtained as a limit.
Let $K\ge1$, $0=m_0\le m_1\le\cdots\le m_K=1$ and define
\begin{eqnarray}
&&q^{r,s}_k(u)=q^{r,s}_k[0]\\
&&q^{r,s}_k(u)=q^{r,s}_k[\ell+1]\;\;\;\mbox{for}\;\;\;m_\ell\le u<m_{\ell+1}\\
&&q^{r,s}_k(1)=q^{r,s}_k[K+1],
\end{eqnarray}
where 
$q^{r,s}_k[\ell]$, $\ell=0,\cdots,K$ are parameters 
satisfying the analog of conditions (\ref{c_simm})-(\ref{c_pd}):
$$
q^{1,1}_k[\ell]=q^{2,2}_k[\ell]\equiv q_k[\ell],\;\;\;
q^{1,2}_k[\ell]=q^{2,1}_k[\ell]\equiv p_k[\ell],
$$ 
$$
q^{r,s}_k[0]=0,\;\;\; q_k[K+1]=1,\;\;\; p_k[K+1]=\tp_k
$$
and
\begin{equation}
\label{pd}
\{q^{r,s}_k[\ell]-q^{r,s}_k[\ell-1]\}_{r,s} \;\;\;\mbox{non-negative definite}.
\end{equation}

If $\alpha=(\alpha_1,\cdots,\alpha_K)$ with 
$\alpha_\ell\in {\mathbb N}$, 
we define the cavity fields as
\begin{eqnarray}
\kappa^{\alpha,s}=y^{(0,s)}+y^{(1,s)}_{\alpha_1}+\cdots+
y^{(K,s)}_{\alpha_1,\cdots,\alpha_K}
\end{eqnarray}
and
\begin{eqnarray}
h_i^{\alpha,s}=z^{(i,0,s)}+z^{(i,1,s)}_{\alpha_1}+\cdots+
z^{(i,K,s)}_{\alpha_1,\cdots,\alpha_K},
\end{eqnarray}
where the $y$'s and the $z$'s are centered Gaussian variables with covariances
\begin{eqnarray}
E (y^{(\ell,s)}_{\alpha_1,\cdots,\alpha_\ell}y^{(\ell',r)}_{\beta_1,\cdots,\beta_\ell'})
&=&\delta_{\ell \ell'}\delta_{\alpha \beta}\frac{\delta^{2d}}{2\gamma^d}\sum_{k,m}
\phi(\delta|k-m|)\left\{q_k^{r,s}[\ell+1]q_m^{r,s}[\ell+1]-q_k^{r,s}[\ell]q_m^{r,s}[\ell]\right\}
\\\nonumber
&=&\frac{\delta_{\ell \ell'}\delta_{\alpha \beta}}2\int_{V\times V}d^d x\,d^dy\,\phi(|x-y|)
\left\{q^{r,s}_x[\ell+1]q^{r,s}_y[\ell+1]-q^{r,s}_x[\ell]q^{r,s}_y[\ell]\right\}
\end{eqnarray}
and
\begin{eqnarray}
E (z^{(i,\ell,s)}_{\alpha_1,\cdots,\alpha_\ell}
z^{(j,\ell',r)}_{\beta_1,\cdots,\beta_\ell'})
&=&\delta_{ij}\delta_{\ell \ell'}\delta_{\alpha \beta}\sum_{m} \delta^d 
\phi(\delta|k-m|)\left\{q_m^{r,s}[\ell+1]-q_m^{r,s}[\ell]\right\}\\\nonumber
&=&
\delta_{ij}\delta_{\ell \ell'}\delta_{\alpha \beta}\left\{\hat q^{r,s}_x[\ell+1]-\hat q^{r,s}_x[\ell]\right\},
\end{eqnarray}
for $i\in \Omega_k$. This is equivalent to say that the parameters $q_k^{\alpha,s;\beta,r}$ in (\ref{h})-(\ref{k})
are given by
\begin{equation}
q_k^{\alpha,s;\beta,r} = q_k^{r,s}[\ell] \;\;\;\mbox{if}\;\;\;\alpha_a=\beta_a\;\;\; \mbox{for}
\;\;\;a<\ell\;\;\;\mbox{and}\;\;\; \alpha_\ell\ne\beta_\ell.
\end{equation} 

Note that the covariance matrix of the $z$'s is 
well defined thanks to condition (\ref{pd}). As for the variables $y$'s,
the same is true thanks to Lemma \ref{lemmino} in Appendix.
As for the random weights $\xi_\alpha$, we choose them as in \cite{ass} to be the Ruelle
probability Cascade with $K$ levels and parameters $m_1,\cdots,m_K$.

At this point, using the properties of the Ruelle Probability Cascades (see for 
instance \cite{T}),
it is not difficult to compute explicitly $\mathcal F_0$.
For the denominator, one finds
\begin{eqnarray}
\gamma^d E\log \sum_\alpha
 \xi_\alpha e^{\beta \sum_s \kappa^{\alpha,s}}=\frac{\beta^2\delta^{2d}}2
\sum_{k,m}\phi(\delta|k-m|)\left[
1+\tp_k \tp_m-\int_0^1du\, \left[q_k(u) q_m(u)+ p_k(u) p_m(u)\right]\right]
\end{eqnarray}
which, in the continuum notation, can be rewritten as
\begin{eqnarray}
\label{polynomial}
&&\frac{\beta^2}{2}\int_V d^d x\left(1+\tp_x^2-\int_0^1 du\,
\left(q_x(u)^2+p_x(u)^2\right)\right)\\\nonumber
&&-\frac{\beta^2}{4}\int_{V\times V}d^dx\, d^dy \,\phi(|x-y|)
\left((\tp_x-\tp_y)^2-\int_0^1 du\,[(q_x(u)-q_y(u))^2+(p_x(u)-p_y(u))^2]
\right)\\\nonumber
&&
+O(\delta)|V|.
\end{eqnarray}

As for the numerator of $-\beta \mathcal F_0$, 
it is not difficult to see that it is bounded above by
\begin{multline}
|\Lambda|2\log2-\gamma^{-d}\int_V d^d x\, \epsilon_x \tp_x\\
+ E \log \sum_\alpha\xi_\alpha 
\prod_k\prod_{i\in\Omega_k}\left[\cosh(\epsilon_k)\cosh(\beta h^{\alpha,1}_i+\beta h)
\cosh(\beta h^{\alpha,2}_i+\beta h)\right.\\
\left.+\sinh(\epsilon_k)\sinh(\beta h^{\alpha,1}_i+\beta h)
\sinh(\beta h^{\alpha,2}_i+\beta h)\right],
\label{lunga} 
\end{multline}
where $\epsilon_k$ has the meaning of a Lagrange multiplier which implements the
constraint $q_k(\sigma^1,\sigma^2)=\tp_k$. The bound holds for any
choice of $\{\epsilon_k\}$, as it follows from the obvious inequality
\begin{eqnarray}
\sum_{q_k(\sigma^1,\sigma^2)=\tp_k}e^{\cdots}=
\sum_{q_k(\sigma^1,\sigma^2)=\tp_k}e^{\cdots+\epsilon_k(q_k(\sigma^1,\sigma^2)-\tp_k)}
\le \sum_{\sigma^1,\sigma^2}e^{\cdots+\epsilon_k(q_k(\sigma^1,\sigma^2)-\tp_k)}.
\end{eqnarray}
Again  using the properties of the Ruelle Cascades,
the expression (\ref{lunga}) can be rewritten as
\begin{eqnarray}
\label{corta}
(2\log 2+O(\delta))|\Lambda|+ 
\gamma^{-d}\int_V d^d x \left(-\epsilon_x \tp_x+\log \cosh(\epsilon_x)+g_x(0,h,h;\epsilon_x)\right),
\end{eqnarray}
where
$g_x(u,h_1,h_2;\epsilon_x)$ 
is the solution of the backward parabolic equation (\ref{eq:antipara}),
with final conditions (\ref{final}).
\hfill $\Box$

At least when $p$ is even, Theorem \ref{th:main} can be immediately extended to the $p$-spin case.

\section{Analysis of the rate function}

The functionals (\ref{fgdrep2}) and (\ref{fgdreppspin}) allow in
principle to study the spin glass problem for small but finite 
$\gamma$.  The difficulties one has to face in this respect are of two types.
First of all, the analysis of the Kac functionals themselves is technically very involved,
since for a given overlap profile one needs to consider a variational principle, whose solution
cannot in general be found explicitly. On the other hand, the analysis of the Kac functional is not
sufficient to infer the behavior of the probability of overlap profiles $P_\Lambda[\{\tp_k\};J]$. 
Indeed, a careful analysis of the error terms due to finite-volume and finite-$\gamma$ effects
is also necessary, as it is already in the ferromagnetic case \cite{cp}.

In order to overcome the first problem, in the present section we restrict to a situation
(temperature close to the critical temperature of mean field theory
and overlap profiles sufficiently smooth to apply a gradient expansion) where the Kac functionals can be 
reasonably replaced by an approximate form, which allows for an explicit treatment of the 
optimization problem. 
The mathematical justification of the approximations involved, together with the study of 
the error terms, is left to future research.

\subsection{Homogeneous solution}

\label{sez:hom}

In order to find the optimal estimate for the free energy $F_\Lambda[\{\tp_k\}]$ of the two-replica
system with constrained overlap profile, one has to minimize the Kac functional
(\ref{fgdrep2}) with respect to 
the Parisi functions $q^{r,s}_x(.)$ and to the Lagrange multipliers $\epsilon_x$.
The variational problem for coupled replicas has an evident
degeneracy, at least in the case of overlap profiles constant in space.
In fact, suppose to take all the $\tp_k$ equal to a value $\tilde{p}$
in the support of the $P(q)$ function for the single-replica problem.
Then, if the Parisi function $q_F(u)$ solves the variational problem in
the case of a single replica, i.e., if it minimizes the expression (\ref{parisi1r}), 
the following position-independent 
form of the functions $q_x(u)$ and $p_x(u)$ solves the two-replica variational problem:
\begin{eqnarray}
& &q_x(u)=\left\{ 
\begin{array}{cc}
q_F(2 u) &\;\;\; u\leq {\tilde{u}}/2\\
\tilde{p}& \;\;\;  {\tilde{u}}/{2}< u\leq \tilde{u}\\
q_F(u) &\;\;\; u> {\tilde{u}}
\label{q2}
\end{array}
\right.\\
& &p_x(u)=\left\{ 
\begin{array}{cc}
q_F(2 u) &\;\;\; u\leq {\tilde{u}}/{2}\\
\tilde{p}&\;\;\;  u> {\tilde{u}}/{2}\\
\end{array}
\right.,
\label{p2}
\end{eqnarray}
where $\tilde{u}$ is defined as the value of $u$ for which
$q_F(\tilde{u})=\tilde{p}$. (It is immediate to verify that the definite positiveness condition
(\ref{c_pd}) is satisfied). At the same time, the variational equations
with respect to  $\epsilon_x$ are solved by $\epsilon_x=0$. It is then possible to see, comparing 
formulas (\ref{fgdrep2}) and (\ref{parisi1r}), 
that with this choice for the variational parameters $q^{r,s}_x$ and $\epsilon_x$, the Kac functional equals
 twice the free energy of
the single-replica mean field system, and is therefore {\em independent of $\tp$}. 
 Indeed, it is immediate to 
 verify this property explicitly for the polynomial part (\ref{polynomial}) of the
 free energy,
which reduces to 
\begin{eqnarray}
-\frac{\beta^2}2|V|\left(1-\int_0^1 du\,q_F(u)^2\right).
\end{eqnarray}
As for the term (\ref{corta}) involving the parabolic equation, 
 one can see that, if  $f_F(u,y)$
is the solution of (\ref{eq:antipara1}) with $q(u)=q_F(u)$, 
 the following form solves the 
 equation (\ref{eq:antipara}) with the choice (\ref{q2})-(\ref{p2}):
 \begin{eqnarray}
 & &g(u,y_1,y_2)=\left\{ 
 \begin{array}{cc}
 2f_F(2 u,y_1)\delta(y_1-y_2) &\;\;\; u\leq {\tilde{u}}/{2}\\
 2f_F(\tilde{u},y_1)\delta(y_1-y_2) & \;\;\;  {\tilde{u}}/{2}\le u\leq \tilde{u}\\
  f_F(u,y_1)+f_F(u,y_2)&\;\;\; u> {\tilde{u}}
 \end{array}
 \right..
 \end{eqnarray}
 Similar considerations   show also that (\ref{corta}) then equals twice the 
term one has in the free energy functional of the SK model.

Notice that in the spin glass $p=2$ case, where at low
temperature the support 
of the function $P(q)$ is an interval $[q_{min},q_{max}]$, this degeneracy  corresponds to the fact that 
the free energy of the two-replica system has
 a continuous zero mode (at least if finite-$\gamma$ effects are neglected). In the $p$-spin case, on the other hand,
the support at low temperature is given by two points
$\{q_0,q_1\}$, and one has just a discrete degeneracy. 

In both cases the degeneracy reflects the fact that a constant overlap
overlap profile, whose value lies in the support of the mean field
distribution function $P(q)$, cannot have too small a probability in
typical samples, as it already follows from the results of
\cite{KacII}.  More precisely: the probability under consideration
could be in principle exponentially small with  the system
size\footnote{ This does certainly happen for instance in dimension
$1$, where there is no phase transition and the overlap is peaked at a
single value as long as $\gamma$ is finite.}, $P\sim \exp(-a(\gamma)
L^{d})$, 
but in that case $a(\gamma)$ has to vanish in the Kac limit $\gamma\to0$.

\subsection{The Parisi model close to $T_c$}

 Since the analysis of the complete functionals is
technically very involved, already at the mean field level it is
customary to resort to simplifying approximations that, while
representing correctly the physics in some limiting cases, allow for
an analytic study of the saddle point equations. In the case of 
$p=2$, where the mean field spin glass transition is of second order,
close to $T=T_c=1$, $h=0$ 
one can use a Landau expansion in the order parameter, 
assuming that the Parisi function $q_F(u)$ is close to zero for any $u<1$. 
On the other hand for the $p$-spin case ($p>2$), where the transition has
a first order character with a discontinuous order parameter, one usually
resorts to a spherical approximation of continuous spins with a global
constraint.

The scope of this section is to begin the study of the large
deviation functional for the Kac spin glass with pair interactions
close to the mean field critical temperature. 
For the SK model,
the Landau expansion truncated to the fourth order is a good
representation of  the free energy of the model in the vicinity if
the critical temperature,  accurate  to the fifth
order in $\tau=T_c-T$.  Parisi proposed \cite{parisimodel} to simplify the expansion,
retaining among all the fourth order terms only the one responsible
for replica symmetry breaking. It has been shown that this
approximation does not affect the free energy to the fifth order in $\tau$.
In the Kac model the Landau expansion can be supplemented by a
gradient expansion, which corresponds to the assumption  that all the functions appearing in
(\ref{fgdrep2}) have small variations in space on scales comparable to
the interaction range $\gamma^{-1}$. The resulting free energy can be
written as: 
\begin{eqnarray}
-\beta  \gamma^d F^{Kac}_\Lambda&=& -\frac{c^*}{2}\int d^dx\;
(\nabla \tp_x)^2+ \frac{\beta^2}{2}\int d^dx\;\tp_x^2+ \frac{1}{2}\int
d^dx\; {\epsilon_x^2} -\int
d^dx\; \tp_x\epsilon_x\nonumber\\
&&-\frac{c}{2} \int dx\; 
[(\nabla \epsilon_x)^2 -\int_0^1 du\; ((\nabla \hat{q}_x(u))^2+(\nabla \hat{p}_x(u))^2)]
\nonumber\\
&&+ \frac{1-T^2}{2} \int dx\; 
[\epsilon_x^2 -\int_0^1 du\; (\hat{q}_x(u)^2+\hat{p}_x(u)^2)]
\nonumber\\
&&+\int d^dx\; \left( \frac{1}{3}
\big[
2\langle \hat{q}_x\rangle\langle \hat{q}_x^2\rangle+\int_0^1 du\; \hat{q}_x(u)\int_0^u dv
(\hat{q}_x(u)
-\hat{q}_x(v))^2+6\langle \hat{p}_x\hat{q}_x\rangle
(\langle \hat{p}_x\rangle -\epsilon_x)\right. 
\nonumber
\\
&&+3\int_0^1 du\, \hat{q}_x(u)\int_0^u dv\;(\hat{p}_x(u)-\hat{p}_x(v))^2
\big]
\nonumber\\
&& \left.+\frac{1}{6} 
[\epsilon_x^4 -\int_0^1 du\; (\hat{q}_x(u)^4+\hat{p}_x(u)^4)]\right)
\nonumber\\
&&
+h^2 \int d^dx\; 
[\epsilon_x -\int_0^1 du\; (\hat{q}_x(u)+\hat{p}_x(u))]
\label{expan}
\end{eqnarray}
where $c=\frac{T^2}{2}\int d^dx\, \phi^{-1}(|x|) x^2$,
$c^*=\beta^4 c$ and we
have introduced the notation $\langle\cdot\rangle=\int_0^1 du\;
\cdot(u)$. 
In performing the expansion, we have assumed that $q_x(u)$ and $p_x(u)$ are of 
order $\tau$ and we have subtracted an inessential constant factor, corresponding to the 
unconstrained free energy of the mean-field single-replica model.

 Instead of trying to justify the expansion (\ref{expan})
we will take it as a {\it fait accompli} and use it for our
preliminary study of the rate function. Of course a full justification
would involve the proof that the neglected terms are much smaller than the
retained one. We expect this to be harmlessly true for $T>T_c$, while 
care should be taken in the discussion of the case $T<T_c$.

\subsubsection{$T>T_c$}

In this section we discuss the interface problem for temperatures
higher than $T_c$, i.e., for $\tau<0$. In these conditions the mean field system is paramagnetic and
for $\gamma\to 0$ the only values of $\tp_k$ giving the same free energy
as the unconstrained solution are given by $\tp_k=0$, and
correspondingly $q_k(u)=p_k(u)=0$. We would like now to impose an overlap value
$\tp_k=\tp_0$ in the boundary and study the decay of the overlap to zero
in the bulk, and the free energy cost for the boundary conditions.  For
simplicity we will suppose that the overlap boundary conditions are
imposed on a given $d-1$ dimensional hyperplane, so that we can limit ourselves
to a one-dimensional problem for the transverse direction. In other
words, the boundary conditions will be $\tp_k=\tp_0$ for all $k$ such that
$k_1=0$, and periodic  in the other
directions. We will look for replica symmetric solutions to the saddle
point equations (which amounts to assume that $q_k(u)=q_k$ and $p_k(u)=p_k$ 
are constant for $0\le u<1$) and use a continuum
limit formulation. The reasons for the replica-symmetric choice 
(apart from its simplicity) will be briefly discussed in the following.
The one-dimensional free-energy density one obtains from (\ref{expan}) is then
\begin{eqnarray}
-\beta \gamma^d f^{Kac}_x&=& \frac{-c^*}{2} (\partial_x \tp_x)^2+
\frac{\beta^2}{2}\tp_x^2+ \frac{1}{2}\epsilon_x^2 -
\tp_x\epsilon_x-\frac{c }{2}[ (\partial_x \epsilon_x)^2
-(\partial_x \hat{q}_x)^2 -(\partial_x \hat{p}_x)^2] \nonumber\\ &&+\tau
(\epsilon_x^2-\hat{q}_x^2-\hat{p}_x^2)+\frac13[2\hat q_x^3+6\hat p_x\hat q_x(\hat p_x-\epsilon_x)]
\nonumber\\
&&+\frac{1}{6} [\epsilon_x^4-\hat{q}_x^4-\hat{p}_x^4]+
h^2(\epsilon_x-\hat{p}_x-\hat{q}_x).
\end{eqnarray}
``Equations of motion''  are obtained by requiring the action to be stationary:
\begin{eqnarray}
&& c^* \partial^2_x \tp_x+\beta^2 \tp_x -\epsilon_x=0\nonumber\\
&&c\partial^2_x \epsilon_x+\epsilon_x-\tp_x+2\tau \epsilon_x-2\hat p_x\hat q_x+\frac23 \epsilon_x^3+
h^2=0
\nonumber\\
&&c\partial^2_x \hat q_x+2\tau\hat q_x-2\hat p_x^2+2 \hat p_x\epsilon_x+\frac23\hat q_x^3+
h^2=0\nonumber\\
&&c\partial^2_x \hat p_x +2\tau \hat p_x-4 \hat q_x\hat p_x+2\hat q_x\epsilon_x+
\frac23 \hat p_x^3+h^2=0.
\end{eqnarray}
The  analysis of these equations for negative $\tau$ is
straightforward. As an illustration we discuss the case $h=0$, the
more general case being quite similar.
For $h=0$ the equations admit a solution $\hat q_x=\hat p_x=0$ and 
\begin{eqnarray}
\nonumber && c^* \partial^2_x \tp_x+\beta^2 \tp_x -\epsilon_x=0\nonumber\\ 
&&c\partial^2_x \epsilon_x+\epsilon_x-\tp_x+2 \tau\epsilon_x+\frac23
\epsilon_x^3=0
\label{equaz}
\end{eqnarray}
Since we are assuming that 
$\tp_x=O(\tau)$, the cubic term in (\ref{equaz}) can be
neglected to the leading order, and one has the solution
\begin{eqnarray}
 \tp_x=\epsilon_x=\tp_0 e^{-\sqrt{2|\tau|/c}\,x}
\end{eqnarray}
which shows that the memory of the boundary condition is lost
exponentially fast. Note that the solution is actually slowly varying, which means 
that the gradient expansion is self-consistent.
Other solutions of (\ref{equaz}) exist (the boundary conditions for $\epsilon_x$ are not
specified) but they are no longer slowly varying or diverge for $x\to\infty$. These have to be discarded, 
since in that case the approximations involved in (\ref{expan}) clearly break down.

Inserting it in the free energy and integrating
over space one finds 
\begin{equation}
-\beta \gamma^d F^{Kac}_\Lambda[\tp_0]= -\sqrt{2c|\tau|}\tp_0^2 (\gamma L)^{d-1}+O(\tilde p_0^4/\sqrt{|\tau|})
(\gamma L)^{d-1},
\end{equation}
which implies that the  boundary conditions imposed require
a surface free energy cost 
\begin{equation}
\label{deltaF}
\Delta F\simeq\frac{L^{d-1}}{\gamma}{\tp_0^2}\sqrt{2c|\tau|}.
\end{equation}
for the
imposed boundary conditions. 
Recall that this result has been obtained within a replica-symmetric Ansatz for the 
Parisi functions. Thanks to the variational character of Theorem \ref{th:main}, this implies that
the free energy cost for the chosen boundary condition is {\sl not smaller} than (\ref{deltaF}).
Note  the appearance
of a divergent length $\xi\sim |\tau|^{-1/2}$ as the transition is
approached. This is the mean field exponent for the correlation
length, which does not take into account renormalization effects.



\subsubsection{$T<T_c$} 

The case $T<T_c$ is the relevant one to study the possibility of
long-range order of Parisi type. Ideally one would like to prove (or
disprove) the existence of a finite critical dimension $d_{LC}$ above
which long-range order is present for small enough $\gamma$. As we
stressed in the introduction, the issue can be formulated as a problem
of sensitiveness to overlap boundary conditions for coupled-replica
systems. The knowledge of the exact large deviation functional, the
free-energy cost as a function of the overlap profile, would in
principle allow to reduce this problem to the proof of the existence
of a phase transition in a (non-random) field-theoretical model. Our
Theorem \ref{th:main} provides a lower bound to the large deviation
functional, thus if we define $d_{LC}^*$ as the lower critical
dimension for the approximated theory,  one can argue that $d_{LC}\leq
d_{LC}^*$\footnote{Of course, one should also show that the ``error
terms'' $O(\delta)$ in Theorem \ref{th:main} are not dangerous }.  In
order to study the stability of long-range order, one should identify
the relevant excitations around the homogeneous solution
(\ref{q2})-(\ref{p2}), that induce overlaps different from $\tp_0$ in
the center of the system, $\tp_0\in Supp(P(q))$ being the chosen
overlap boundary condition.  In the following of this section we will
argue that excitations localized in a region of size $\ell$ have a
free-energy cost of the order $\ell^{d-a}/\gamma^a$. On the other
hand, in analogy with the usual Peierls argument for the Ising model,
one should also find a way to estimate the number of such excitations
and show that the corresponding entropic gain does not overcome the
free-energy cost at low enough $\gamma$. Let us for example imagine
that the number of excitations grows with $\ell$ as $\exp(c
\ell^{d-b})$. It is clear that if $b=a$, as it happens in models
without disorder, then for $\gamma$ small enough and $d>a$ large
excitations will be exponentially suppressed and
$d_{LC}^*=a$. Conversely, if $b< a$ the entropy terms will dominate
for all $d$ and destroy the phase transition for all finite $\gamma$.

An equivalent method to determine long-range order makes use of ``twisted''
boundary conditions. In that case one direction is singled out and,
while periodic conditions are assumed in the remaining directions,
$\tp_1$ and $\tp_2$ conditions are chosen at the boundaries along the
preferred direction \cite{fpv}. Again, in a system of linear size $L$ one can
expect profiles with free-energy cost proportional to
$L^{d-a}/\gamma^a$ and an entropy proportional to $(\gamma
L)^{d-b}$.

The ``energetic'' contribution can be estimated for small $\gamma$ as
the value of the large-deviation functional corresponding to the
minimizing profile. The study of non-uniform saddle points of the 
free-energy (\ref{expan}) corresponding to inhomogeneous boundary
conditions to determine the cost for overlap interfaces have been
first put forward in \cite{fpv}. In that paper the saddle point
procedure was assumed without further justification, while here it
legitimated by the appearance of the interaction volume $\gamma^{-d}$ in
front of the action. The results of \cite{fpv} can be mutuated
directly to our case.

In that paper it was
assumed that the solution of the optimization problem could be
obtained by a perturbation of the form (\ref{q2})-(\ref{p2}), and it
was found that the form (\ref{q2})-(\ref{p2}) should be modified for
values of $u$ around $\tilde{u}/2$ and $\tilde{u}$.  Thanks to this
modification, and translating in terms of the Kac model, the scaling
of the free energy results to be of the order
$L^{d-5/2}/\gamma^{5/2}$, thus modifying the values of the exponent
$a$ from $3$ to $5/2$. This again should be considered as a (better)
upper bound estimation, and if we suppose that this is the exact
value, we conclude as in \cite{fpv} that ``energetic'' effects destroy
the possibility of Parisi order in dimension $d=1,2$, while they are
consistent with that kind of order in dimension 3 and above.

We would like to conclude by remarking that a simple estimate making use of the
``homogeneous solution'' (\ref{q2})-(\ref{p2}) does not apply to the $p$-spin model, and more
generally to models with one-step RSB. In these cases the function
$q_F(u)$ has the form $q_F(u)=\theta(m-u)q_0+\theta(u-m)q_1$ and the
support of $P(q)$ is just the set  $\{q_0,q_1\}$.
There are only two homogeneous solutions of the kind
(\ref{q2})-(\ref{p2}) corresponding to the two possible costless overlap
values for $\tp$, $q_0$ and $q_1$. The analogous of the hypothesis of
neglecting the kinetic contribution in the saddle point equations
would consist in assuming in different points of space either one or
the other possible form. But in this case, it is simple to realize
that the kinetic term
would be identically equal to zero. In order to get an estimate of the
free-energy cost for an interface, a detailed solution of the space-dependent 
saddle point equations is necessary \cite{BiaFra}.

\section{Conclusions}

The main result of this paper is the derivation of large deviation
functional, quantifying for small $\gamma$ the probability of overlap
profiles in spin glass models with Kac-type interactions. We obtain
that the replica expression provides a lower bound to the true
free-energy functional. Moreover, putting aside the problem of 
mathematically justifying the approximations involved,
we performed a first analysis of the free-energy functional, 
finding estimates for the free-energy cost of extended ``overlap
interfaces''.

In this paper we have considered the case of
two replicas coupled in a symmetric way. Although we did not discuss
it, the generalization of our analysis to $R$ replicas, which can be
useful to discuss issues related to ultrametricity \cite{fpvII}, can
be achieved in a rather straightforward way. Another possible and more
interesting generalization concerns the introduction of a ``quenched
potential'' where again one considers two replicas, but coupled in an
asymmetric way. One considers a first replica in an arbitrary
configuration chosen with the usual Boltzmann weight, and then a
second one, constrained to have a given overlap profile with the first
\cite{fp}. This approach,  more involved than the one
presented here, would be especially relevant for the case of the
$p$-spin model, where one would like to study nucleation of entropic
droplets in a given equilibrium state \cite{wol,birbou}. 

\section*{Acknowledgments} 
We thank T. Bodineau, M. Cassandro and J. Kurchan for interesting discussions. 
This work was supported in part by the European Community's 
``Human Potential Programme'' under contract  HPRN-CT-2002-00319 STIPCO  
and by the Swiss Science Foundation Contract No. 20-100536/1.

\appendix

\section{Covariance of the cavity fields $y^{(\ell,s)}_{\alpha}$}

Let $\ell=0,\cdots,K$, $r=1,2$, and 
\begin{eqnarray}
\hat A^{(r)}_\ell=\left(
\begin{array}{cc}
a_{11}^{(r)}(\ell) &a_{12}^{(r)}(\ell)\\
a_{12}^{(r)}(\ell) &a_{11}^{(r)}(\ell)\\
\end{array}
\right).
\end{eqnarray}
The positive-definiteness of the covariance of the variables $y^{(\ell,s)}_{\alpha}$
in Section \ref{sec:int_est} follows from the following Lemma:
\begin{lemma}
\label{lemmino}
Assume that $\hat A^{(r)}_\ell-\hat A^{(r)}_{\ell-1}$, $\ell=1,\cdots,K$, $r=1,2$
are non-negative definite. Then, the same holds for $\hat B_\ell-\hat B_{\ell-1}$, where
\begin{eqnarray}
\hat B_\ell=\left(
\begin{array}{cc}
a_{11}^{(1)}(\ell)a_{11}^{(2)}(\ell) & a_{12}^{(1)}(\ell)a_{12b}^{(2)}(\ell) \\
a_{12}^{(1)}(\ell)a_{12}^{(2)}(\ell) & a_{11}^{(1)}(\ell)a_{11}^{(2)}(\ell) \\
\end{array}
\right)
\end{eqnarray}
\end{lemma}
{\em Proof of Lemma \ref{lemmino}}  From the hypothesis follows that
\begin{eqnarray}
\label{pos1}
&&a_{11}^{(r)}(\ell)\pm a_{12}^{(r)}(\ell)\ge0 \\
\label{pos2}
&&\Delta_{11}^{(r)}\pm \Delta_{12}^{(r)}\ge0,
\end{eqnarray}
where $\Delta_{cd}^{(r)}=a_{cd}^{(r)}(\ell)-a_{cd}^{(r)}(\ell-1).$
The statement of the Lemma is equivalent to the non-negativity of 
\begin{eqnarray}
\label{aux}
[a_{11}^{(1)}(\ell)a_{11}^{(2)}(\ell)-a_{11}^{(1)}(\ell-1)a_{11}^{(2)}(\ell-1)]\pm 
[a_{12}^{(1)}(\ell)a_{12}^{(2)}(\ell)-a_{12}^{(1)}(\ell-1)a_{12}^{(2)}(\ell-1)],
\end{eqnarray}
which follows immediately from Eqs. (\ref{pos1})-(\ref{pos2}),
once (\ref{aux}) is rewritten as
\begin{eqnarray}
&&(\Delta_{11}^{(1)}+\Delta_{12}^{(1)})\frac{a_{11}^{(2)}(\ell)\pm a_{12}^{(2)}(\ell)}2
+(\Delta_{11}^{(1)}-\Delta_{12}^{(1)})\frac{a_{11}^{(2)}(\ell)\mp a_{12}^{(2)}(\ell)}2
\\\nonumber
&&+(\Delta_{11}^{(2)}+\Delta_{12}^{(2)})\frac{a_{11}^{(1)}(\ell-1)\pm a_{12}^{(1)}
(\ell-1)}2
+(\Delta_{11}^{(2)}-\Delta_{12}^{(2)})\frac{a_{11}^{(1)}(\ell-1)\mp a_{12}^{(1)}(\ell-1)}2.
\end{eqnarray}
\hfill $\Box$

\end{document}